\pdfoutput=1

\documentclass[11pt,a4paper]{article}
\usepackage[hyperref]{acl2020}
\usepackage{times}
\usepackage{latexsym}

\usepackage{microtype}

\aclfinalcopy 

\usepackage{amsmath,amsfonts,bm}

\def\eqref#1{equation~\ref{#1}}

\def\1{\bm{1}}

\DeclareMathAlphabet{\mathsfit}{\encodingdefault}{\sfdefault}{m}{sl}
\SetMathAlphabet{\mathsfit}{bold}{\encodingdefault}{\sfdefault}{bx}{n}

\usepackage{color}

\usepackage[utf8]{inputenc} 
\usepackage[T1]{fontenc}    
\usepackage{hyperref}       
\usepackage{url}            
\usepackage{booktabs}       
\usepackage{amsfonts}       
\usepackage{nicefrac}       
\usepackage{microtype}      
\usepackage{graphicx}
\usepackage{multirow}
\usepackage[inline]{enumitem}
\usepackage{subcaption}

\usepackage{tabularx}
\usepackage{blindtext}

\newcolumntype{L}[1]{>{\raggedright\arraybackslash}p{#1}}
\newcolumntype{C}[1]{>{\centering\arraybackslash}p{#1}}
\newcolumntype{R}[1]{>{\raggedleft\arraybackslash}p{#1}}

\title{Braille-to-Speech Generator: Audio Generation Based on Joint Fine-Tuning of CLIP and Fastspeech2}

\author{%
  Chun Xu\thanks{\hspace{.3em} Equal contribution} \quad 
  En-Wei Sun \\ 
  Xinjiang University of Finance and Economics \\
  \texttt{2022210236@xjufe.edu.cn} \\
}

\begin{document}

\maketitle

\begin{abstract}
An increasing number of Chinese people are troubled by different degrees of visual impairment, which has made the modal conversion between a single image or video frame in the visual field and the audio expressing the same information a research hotspot. Deep learning technologies such as OCR+Vocoder and Im2Wav enable English audio synthesis or image-to-sound matching in a self-supervised manner. However, the audio data used for training is limited and English is not universal for visually impaired people with different educational levels. Therefore, for the sake of solving the problems of data volume and language applicability to improve the reading efficiency of visually impaired people, a set of image-to-speech framework CLIP-KNN-Fastspeech2 based on the Chinese context was constructed. The framework integrates multiple basic models and adopts the strategy of independent pre-training and joint fine-tuning. First, the Chinese CLIP and Fastspeech2 text-to-speech models were pre-trained on two public datasets, MUGE and Baker, respectively, and their convergence was verified. Subsequently, joint fine-tuning was performed using a self-built Braille image dataset. Experimental results on multiple public datasets such as VGGSound, Flickr8k, ImageHear, and the self-built Braille dataset BIT-DP show that the model has improved objective indicators such as \textit{BLEU4},\textit{FAD}(Fréchet Audio Distance), \textit{WER}(Word Error Ratio), and even inference speed. This verifies that the constructed model still has the ability to synthesize high-quality speech under limited data, and also proves the effectiveness of the joint training strategy that integrates multiple basic models.
\end{abstract}

\begin{figure*}[thb]
  \centering
  \includegraphics[width=.6\linewidth]{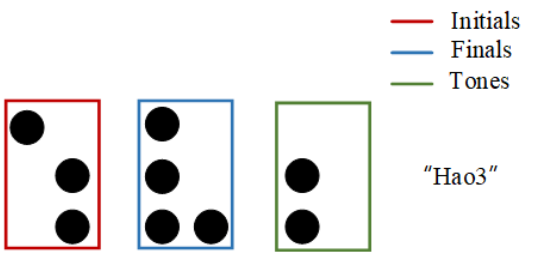}
  \caption{Braille "Hao3" Example ( The first square in red represents the initial consonant "h", the second square in blue in the middle represents the compound vowel "ao", and the third square in green indicates the third tone, i.e., the rising tone.).}
  \label{fig.1}
\end{figure*}

\section{Introduction}
As of now, there are over 17 million individuals with visual impairments in China, and the number of visually impaired individuals of school age has been steadily increasing. The issue of visual impairment poses significant challenges to education. In the process of acquiring knowledge and external stimuli, tactile sensation and audio are undoubtedly the two main avenues for individuals with visual impairments. Among them, physical reading materials and teaching materials mainly use braille with raised dots, which generally consist of three or two dots forming a syllable granularity pinyin rather than Chinese characters~\citep{huang2023translating}. As shown in Figure~\ref{fig.1}. Therefore, compared to regular text materials, braille has the disadvantage of restricting the speed of knowledge acquisition due to its longer length when expressing the same meaning ~\citep{zhu2024towards}.When implementing higher education that encompasses a broader scope of knowledge, the process of manually recording ordinary textbooks into audio books essentially involves treating the human eye as an image collector to gather visual information containing knowledge and transforming it into audio information that is understandable for the visually impaired population. However, this task not only consumes a significant amount of time but also requires a considerable number of experienced personnel familiar with the rules of compiling braille image-to-audio conversion. Nonetheless, relying solely on human expertise does not guarantee efficiency and accuracy in the conversion process.

At present, there are two mainstream deep learning frameworks for achieving image-to-audio conversion. On the one hand, the two-stage task of image-text-speech involves using Optical Character Recognition (OCR) to recognize text in images and convert it into text, followed by using a spectrogram encoder-decoder to convert the text into a spectrogram and decode it into speech. On the other hand, the single-stage task of image-speech involves using a self-supervised approach to learn large-scale image-speech pairs using visual and audio encoders separately~\citep{li2023trocr,reddy2021recognizing,rodriguez2023ocr,wang2023survey}. Images and speech are embedded into a shared representation space to capture the interaction information between the two modalities~\citep{chi2023tiar,2020Perfect,huang2023self,wang2024speechclip+}.

However, OCR requires high-quality image annotations to recognize and output text from images without text. During model training, high-quality audio datasets are often too small to obtain satisfactory model performance. Of course, when using self-built datasets for fine-tuning, the preprocessing of audio data is more complicated than that of text data, including alignment in the time domain~\citep{chen2023dialogmcf}, issues with speech rate~\citep{oh2024diffprosody}, and background noise~\citep{varkonyi2023dynamic},etc.At the same time, since these deep learning models have not been pre-trained on large-scale Chinese data, it is difficult to synthesize high-quality Chinese audio.

To address the aforementioned issues, this paper constructs a two-stage model for image-text-speech conversion, named CLIP-KNN-Fastspeech2, based on a set of base models. In the image-text stage, the Chinese CLIP (Contrastive Language-Image Pre-Training) model is pre-trained using self-supervised learning and contrastive learning to fully learn the features of image-text pairs and make up for the image annotation requirements of OCR. The K-Nearest Neighbor (KNN) model retrieves text information from images based on the feature differences between positive and negative samples. In the text-speech stage, Fastspeech2 takes the text output from the previous stage as input, generates the corresponding mel spectrogram, and decodes it into speech. After pre-training CLIP and Fastspeech2, the model is fine-tuned using a self-constructed braille-pinyin dataset. Since pinyin is composed of initials, finals, and tones arranged in accordance with rules, annotating image-text data is easier to achieve than preprocessing audio data.

\section{Related Work}

The purpose of image-to-text conversion is to obtain corresponding text representations from input images~\citep{zhu2023peit,zhu2024vistfc}. Image-text models generally consist of three modules: a visual encoder to capture fine-grained image features, a text encoder to perceive contextual semantic connections, and a sequence decoder. The decoder outputs corresponding text fields based on hidden features.~\citep{bao2022vl}proposed the VL-BEiT model based on a bidirectional multimodal Transformer for visual-language pre-training. By predicting masks for input image-text pairs, it captures local connections between the two modalities.~\citep{bao2022vlmo} designed a hybrid modality expert Transformer structure that can be used as both a dual encoder for image-text retrieval and a fused encoder to simulate deep interaction between image-text pairs. This architecture uses separate pooling layers and shared self-attention mechanisms for specific modalities.~\citep{yu2022coca} proposed the CoCa (Contrastive Captioner) model, which decomposes the decoding layer into two parts: the first half omits cross-modal attention to encode unimodal text representations, while the remaining decoding layer realizes multimodal image-text representations. The model is trained using contrastive loss between jointly embedded image-text pairs and subtitle loss of the multimodal decoder.~\citep{li2022blip} designed the BLIP (Bootstrapping Language-Image Pre-training) model for visual language understanding, optimizing it from three aspects: image-text contrastive loss, language modeling loss, and image-text matching loss, to achieve more powerful image-text understanding.~\citep{li2023scaling} proposed FLIP (Fast Language-Image Pre-training) based on the CLIP model, which mainly reduces training time and improves accuracy by randomly masking and removing parts of images.~\citep{li2021align} constructed the ALBEF (Align Before Fuse) model based on cross-modal attention, which achieves modality alignment before encoding image-text information and proposes momentum distillation for self-training, learning modality interaction information from generated pseudo-targets.~\citep{chen2023vilem}, in order to enhance the discriminative ability of local text semantics, associated local text semantics with high-level visual context and multi-level local visual information using the ViLEM model.~\citep{yang2022chinese} extended the CLIP model to support the Chinese language scene and pre-trained it on a large-scale Chinese dataset. Experimental results demonstrate its state-of-the-art performance in the Chinese domain.

Text-to-speech aims to synthesize natural and understandable speech given text~\citep{zhao2023ccsrd,lei2023ckdst,sun2023towards}. It first converts text into sequences of phonemes or graphemes with acoustic features (such as linear spectrograms or mel spectrograms), and then the vocoder transforms these acoustic features into corresponding audio samples. SHEN et al.~\citep{shen2018natural} designed Tacotron2, a sequence-to-sequence feature prediction network based on Recurrent Neural Networks (RNNs), which converts character embeddings into mel-scale spectrograms and synthesizes spectrograms into time-domain waveforms using an improved WaveNet model.~\citep{ren2020fastspeech} proposed Fastspeech2, an end-to-end acoustic model that generates speech waveforms directly from text by conditioning on factors such as duration in speech waveforms.~\citep{yang2023diffsound} introduced Diffsound, based on a discrete diffusion model, to overcome biases and error accumulation caused by autoregressive decoding predictions.~\citep{kreuk2022audiogen} developed AUDIOGEN, an autoregressive generative model, contributing to individual voice characterization, background noise and reverberation handling, addressing the problem of long sequences at high sampling rates, and data annotation processing.~\citep{huang2023make} established Make-an-Audio2, a latent diffusion model, to enhance semantic alignment and temporal consistency through structured text input and designing a diffusion denoiser based on feedforward Transformers.~\citep{liu2023audioldm} designed AudioLDM, a model that learns latent representations of audio instead of cross-modal relationship modeling. ~\citep{sofer2024c} proposed C-CLAPA, a subtitle decoder audio pre-training method based on data augmentation.After reducing the loss in cross-modal data conversion between text and speech, more researchers have begun synthesizing audio that better reflects the emotions of the text by designing emotion extractors~\citep{li2024mm,lin2023stylebert,zhang2023deep}.

For implementing image-to-speech conversion using a two-stage approach, it is crucial to utilize or design a bridging module that connects the I2T (Image-to-Text) system and the T2A (Text-to-Audio) system.~\citep{wang2024v2a} proposed the V2A-Mapper to bridge the domain gap between visual CLIP and the auditory model CLAP (Contrastive Language-Audio Pre-training) in latent space, thus enhancing the fidelity of generated audio.~\citep{chen2023end} increased the naturalness of generated audio by adding a duration predictor to predict phoneme durations within characters and reconstructing input words based on factorized durations.

\section{Method}

\begin{figure*}[thb]
  \centering
  \includegraphics[width=.9\textwidth]{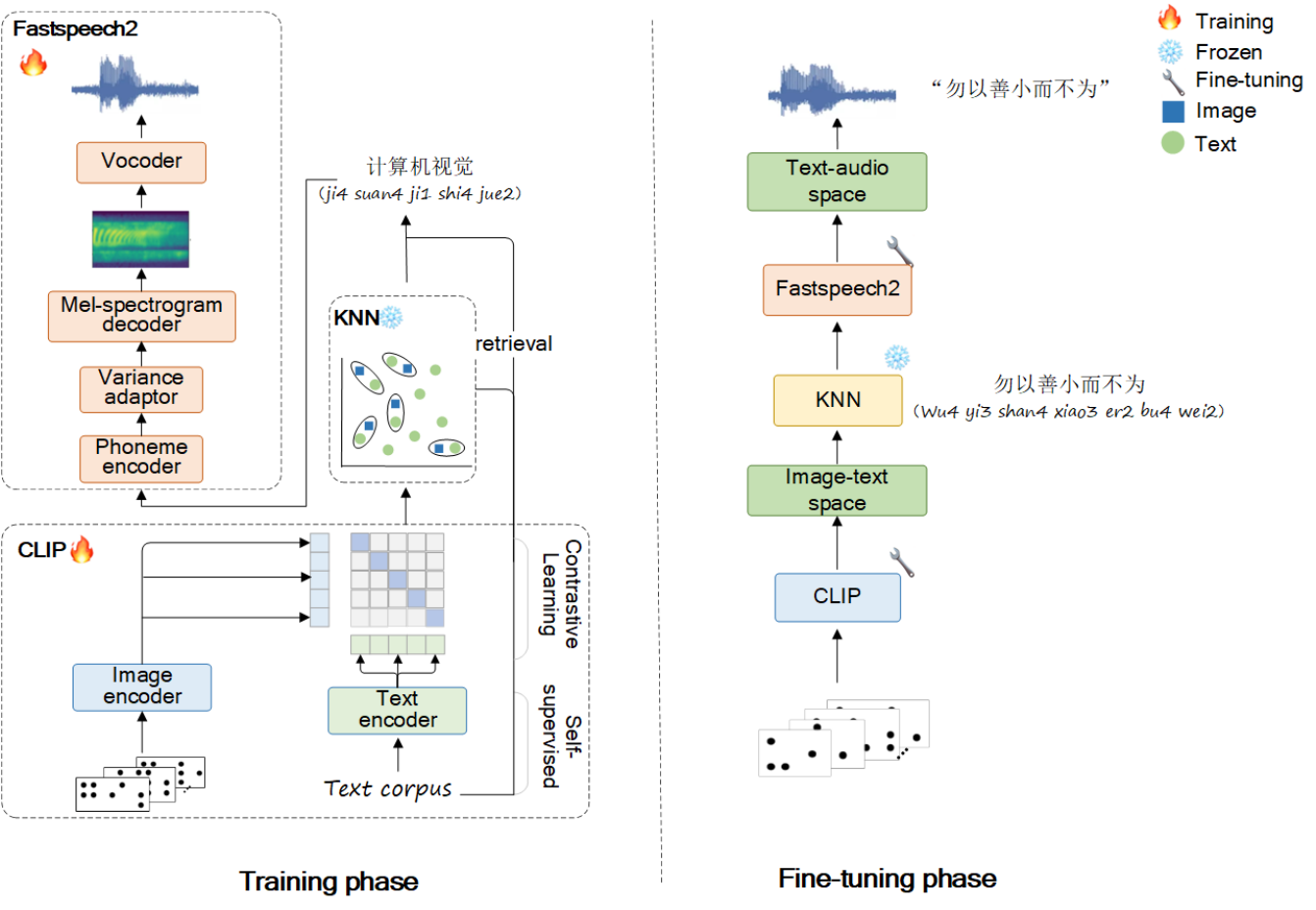}
  \caption{Architecture of the Two-Stage Image-to-Speech Model (Left: In the I2T stage, the pre-trained Chinese CLIP model learns features of image-text pairs and retrieves text through the KNN network; in the T2A stage, the pre-trained Fastspeech2 vocoder demonstrates its audio-text conversion capability. Right: Fine-tuning on the braille dataset is conducted using the trained CLIP-KNN-Fastspeech2 model.).}
  \label{fig.2}
\end{figure*}

\subsection{Framework Overview}
From the perspective of data language: In specific Chinese contexts, existing end-to-end image-to-speech systems fail to simultaneously demonstrate outstanding data adaptation effects in both the image-to-text and text-to-audio stages. From the standpoint of data quantity and quality: The modality conversion effects of the I2T (Image to Text) and T2A (Text to Audio) systems are based on a large quantity of high-quality annotated data. Self-supervised learning models for image-to-speech, represented by SCRL (Semantic-Consistent Representation Learning)~\citep{ning2021semantics} and MCRN (Multi-source Cross-modal Retrieval Network)~\citep{yuan2022mcrn}, learn the interaction of information between modalities from datasets comprising millions of audio data points. Therefore, the model for braille image-to-speech conversion constructed in this paper integrates representative and generative basic models learned from large datasets, relaxing the requirements for difficult-to-obtain and process data types when implementing specific downstream tasks.

As shown in Figure~\ref{fig.2}, the framework chosen for the I2T stage only learns features of the image-text modality and combines them with a KNN network for text retrieval rather than an end-to-end image-text model. Firstly, the image encoder and text encoder serialize braille images and corresponding text sequences into feature vectors. The relationship matrix quantifies the matching degree between images and text, and selects non-paired image-text pairs with high similarity as negative samples for contrastive learning. This further enhances the matching degree of paired image-text pairs while reducing the matching degree of non-paired image-text pairs. Then, using the KNN network, braille corresponds to Pinyin, numbers, and other text based on the learned latent features. In the T2A stage, the text output from the previous stage is encoded, and a variable adaptation layer is designed to predict phonemes, tones, and volume to capture audio features. The mel-spectrogram generator generates mel-spectrograms, and finally, the vocoder synthesizes the original waveform based on the synthesized mel-spectrogram.

\textbf{CLIP}\quad In order to better establish the connection between images and text, we chose the CLIP model instead of the traditional YOLO series of visual capturers. Braille images contain fewer information elements than common physical images, such as landscapes, objects, and portraits. CLIP uses image encoders and text encoders to serialize images and texts into feature vectors \textit{$\vec {a}$} and \textit{$\vec {b}$} that interact with each other in the same semantic space, and determines the strength of the match between the image and text by whether the dot product \textit{p}=\textit{$\vec {a}$}·\textit{$\vec {b}$} tends to 1 or 0. This allows the association between images and text to be established after pre-training on a large-scale corpus, so the CLIP model is chosen to avoid the complexity of image annotation.

\textit{Text Encoder}\quad It follows the I2T architecture described in Figure 2, using the Chinese encoder RoBERTa-wwm-ext-large-Chinese~\citep{liu2019roberta} from the CLIP model to dynamically mask the input text for tokenization. It continuously learns the contextual semantic connections of the text using multi-layer attention mechanisms and positional encodings.

\textit{Image Encoder}\quad First, the visual processing model ViT-H/14~\citep{dosovitskiy2020image} (Vision Transformer-Huge/14x14 pixel) divides the braille image into a series of image patches and applies linear mapping to obtain a sequence of vectors. Then, in each Transformer module, self-attention layers and feed-forward neural networks are designed to capture the correlations between image patches and represent the contextual-aware features of the image patches. Finally, global average pooling is applied to integrate the features of the image patch sequence into a global feature vector, which is then classified using linear layers.

\textbf{KNN}\quad After the image-text matching strength \textit{p} is calculated by CLIP, to ensure the integrity and accuracy of the system, we use the KNN method to set the \textit{k} value and sort the k potential texts corresponding to the image by the p value. It is worth noting here that we use the p value instead of the Euclidean distance in KNN.

\textbf{Fastspeech2}\quad Compared with the Tacotron series and Fastspeech, the Fastspeech2 used in this article has improved the sound quality of synthesized audio and the training speed. During the training process, the delay, pitch and energy information of the speech are extracted as conditional input to alleviate the one-to-many mapping problem. This is due to the fact that Fastspeech2 provides more speech change information.

\textit{Variance Adaptor}\quad To train the model to generate high-quality audio outputs, the extracted audio duration, pitch, and energy are used as inputs to the hidden sequence. A duration predictor is designed to predict the duration of each factor, and the prediction is optimized through mean square error loss. The pitch predictor decomposes the continuous high-pitch sequence into pitch spectrograms using continuous wavelet transform for training. The energy predictor calculates the \textit{L2} norm of the amplitude of each short-time Fourier transform (STFT) frame as energy, uniformly decomposes the energy of each frame, and encodes it into an energy vector. The architectures of these three predictors are consistent: they consist of two layers of one-dimensional convolutional layers with ReLU activation function, followed by dimension reduction through normalization layers, and then output through a linear layer after passing through another normalization layer.

\textit{Vocoder}\quad To balance between the speed and quality of speech synthesis, one of the foundational vocoder models, HiFi-GAN~\citep{kong2020hifi} is utilized. It consists of a generator and two discriminators composed of multiple layers of Convolutional Neural Networks (CNNs) and Transpose Convolutional Neural Networks (TCNNs). The generator converts input mel-spectrograms into speech waveforms, while the discriminators, MPD (Multi-period discriminator) and MSD (Multi-scale discriminator), are responsible for identifying signals of different periods in speech and handling excessively long data, respectively.

\subsection{Datasets}
In order to increase the credibility of the proposed model CLIP-KNN-Fastspeech2,the Flickr8k~\citep{harwath2015deep} multi-speaker natural speech dataset ,VGGSound~\citep{chen2020vggsound} and ImageHear~\citep{sheffer2023hear} are used to verify the improvement of the proposed model compared with the existing model.At the same time, we also utilized three Chinese datasets to discuss the generalization and robustness of the model:Wukong~\citep{gu2022wukong} and the self-built braille dataset BIT.

\begin{figure*}[thb]
  \centering
  \includegraphics[width=0.9\textwidth]{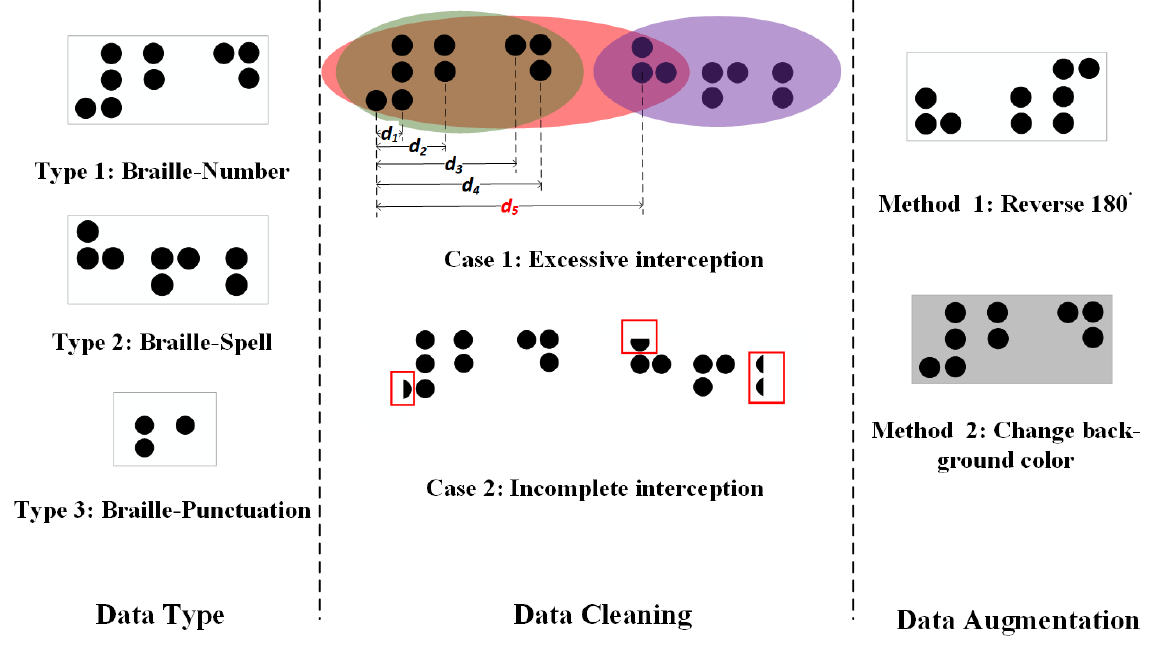}
  \caption{Types, Cleaning, and Enhancement Methods of the BIT Dataset. \textit{Left Figure}: Illustrates the three categories of Braille correspondence contained in the dataset, where Type 1 corresponds to the numeral "24", Type 2 corresponds to the Pinyin "hao4", and Type 3 corresponds to the punctuation mark "!". \textit{Middle Figure}: Presents two cases of "dirty data", namely, Excessive cropping and Incomplete cropping.\textit{Right Figure}: Demonstrates two methods for data augmentation, namely, Flipping the image by 180 degrees and Adding background colors.).}
  \label{fig.3}
\end{figure*}

\textbf{BIT}\quad Due to the possibility of recognition errors caused by the confusion in multi-party screenshot extraction~\footnote{https://http://www.braille.org.cn/} during the conversion from Chinese characters to Braille images in the self-built \textbf{B}raille \textbf{I}mage-\textbf{t}ext dataset, it is necessary to perform data cleaning and augmentation to ensure image quality and model feature extraction capability, as shown in Figure~\ref{fig.3}.

In Figure 3, the green and purple ellipses represent "24" and "hao4" respectively, while the red ellipse indicates the erroneously cropped "248" due to overcropping. To address the semantic confusion caused by overcropping , data cleaning is conducted using absolute distances. The leftmost point serves as a reference to determine the correct distance set, denoted as $G=[d_1,d_2,d_3,d_4]$. The appearance of distance $d_5$ indicates overcropping. For data in Case 2, manual observation is employed for data selection.

Meanwhile, we constructed the BIT dataset, initially comprising 6200 image-text pairs and three data types: Braille-punctuation, Braille-numbers, and Braille-pinyin. After data cleaning and augmentation, the BIT-DP (BIT with \textbf{D}ata \textbf{P}rocessing) dataset was expanded to include 11,200 data instances. The original BIT dataset was subdivided into two sub-datasets to facilitate subsequent ablation experiments: (1) concerning data processing, BIT-DC (BIT with \textbf{D}ata \textbf{C}leaning) and BIT-DA (BIT with \textbf{D}ata \textbf{A}ugmentation); (2) concerning data categories, BIT-N (BIT in \textbf{N}umber) and BIT-S (BIT in \textbf{S}pell). The specific division of the dataset and its rationale are presented in Table~\ref{tab3}.

\begin{table}[thb]
\centering
\caption{BIT dataset division}
\resizebox{1\linewidth}{!}{
\begin{tabular}{cccc}
\toprule
\textbf{Datasets} & \textbf{Braille-Number} & \textbf{Braille-Spell} & \textbf{Braille-Punctuation} \\ \midrule
BIT    &  2000  & 2700  & 1500 \\
BIT-DC &1800  & 2500  & 1300 \\
BIT-DA &4000  & 5400  & 3000 \\
BIT-DP &3600  & 5000  & 2600 \\ \midrule
BIT-N &3600  & -     & - \\
BIT-S &-     & 5000  & - \\
BIT-P &-     & -     & 2600 \\ \bottomrule
\end{tabular}
}
\label{tab3}
\end{table}

\subsection{Training}
To address the issue of computational load on devices, the model will adopt the following multi-step training approach. Following the CLIP~\footnote{https://github.com/OFA-Sys/Chinese-CLIP} contrastive learning philosophy, the model will balance the feature extraction performance for both images and texts by averaging the losses of image embeddings and text embeddings. The losses for image embeddings and text embeddings are computed using cross-entropy functions to measure the discrepancies between the model's outputs and the ground truth. AS equations (1) to (3) shown:
\begin{equation}
  \ Loss_i=CE\_loss(label_{output(i)},label_i)
\end{equation}
\begin{equation}
  \ Loss_t=CE\_loss(label_{output(t)},label_t)
\end{equation}
\begin{equation}
  \ Loss_{i-t}=(Loss_i+Loss_t)/2
\end{equation}
where $label_{output(i)}$ represents the predicted labels for the image outputs of the model, while $label_i$ represents the true labels for the images. Similarly, $label_{output(t)}$ and $label_t$ denote the predicted text labels and the true text labels, respectively.

After then, the pretrained image and text encoder parameters are frozen, and the training of the Fastspeech2~\footnote{https://github.com/ming024/FastSpeech2} vocoder begins. The training process mainly consists of two major modules: mel-spectrum prediction and variable adapter parameter tuning. The mel-spectrum prediction loss consists of spectrum generation loss and spectrum encoding loss. The former is primarily used to distinguish the differences between the mel-spectrogram generated by the model and the ground truth mel-spectrogram, while the latter measures the model's ability to map each character in the input text to the corresponding time step (or frame) in the mel-spectrogram. Shown as the equations (4) and (5) for specifics:
\begin{equation}
  \ Loss_{mel}=MSE\_loss(mel,mel_{target})
\end{equation}
\begin{equation}
  \ Loss_{mel}'=MSE\_loss(mel_{post},mel_{target})
\end{equation}
Where the variables $mel$, $mel_{post}$, and $mel_{target}$ are all new tensors obtained after masked selection, containing only the elements of the original tensor corresponding to True values in the mask. This is done to match the length of the text sequence and determine whether the corresponding spectrogram positions are valid, dynamically adjusting the length of the generated mel-spectrogram.

The training of the variable adapter involves the prediction loss for duration, pitch, and energy. Specifically, the duration loss and pitch loss are mapped to exponential and logarithmic functions, respectively, to simulate subjective perception. Shown as equations (6) to (8) for details:
\begin{equation}
  \ Loss_{duration}=MAE\_loss(d_{pre},d_{target})
\end{equation}
\begin{equation}
  \ Loss_{pitch}=MAE\_loss(p_{pre},p_{target})
\end{equation}
\begin{equation}
  \ Loss_{energy}=MAE\_loss(e_{pre},e_{target})
\end{equation}
Where $Loss_{duration}$  is measured by mean squared error to quantify the difference between $d_{pre}$ and $d_{target}$. Unlike $p_{pre}$, $e_{pre}$, and others, which form new tensors by removing blank or padding symbols from the mel spectrogram mask, $d_{pre}$ and $d_{target}$ are new tensors composed of valid characters based on the source input sequence mask. Hence, the overall objective of training the Fastspeech2 vocoder is $Loss_{t-a}$, as specified in equation (9):
\begin{equation}
\begin{split}
 \ Loss_{t-a} &= Loss_{mel}+Loss_{mel}'+Loss_{duration} \\
              &+Loss_{pitch}+Loss_{energy}
\end{split}
\end{equation}

After separately completing the pre-training stages for both the I2T and T2A phases, the model is fine-tuned using self-constructed Braille data, and the two-stage training losses are summed to obtain the parameter optimization direction for the CLIP-KNN-Fastspeech2 model, denoted as $Loss_{total}$, as shown in equation (10):
\begin{equation}
  \ Loss_{total}=\lambda_{1}Loss_{i-t}+\lambda_{2}Loss_{t-a}
\end{equation}
Where $\lambda_{1}$ and $\lambda_{2}$ are custom weight parameters, with $\lambda_{1}, \lambda_{2}\in \left ( 0 ,1\right) $, and the sum of $\lambda_{1}$ and $\lambda_{2}$ is equal to 1.Here we consider the equal importance of the two phases, preliminarily set the parameter $\lambda_{1}$ = $\lambda_{2}$ = 0.5, and the influence of different parameter design on the model will be discussed in the following section 5.3.

\subsection{Parameter Settings and Pre-training Process}
\textbf{CLIP}\quad Pretraining of CLIP's visual encoder and text encoder was conducted using the publicly available dataset MUGE~\citep{lin2021m6}. The MUGE dataset comprises a training set with a total of 250,000 image-text pairs, while both the test set and validation set consist of 5,000 search IDs each. Retrieval is performed from respective candidate pools of 30,000 images, and the data format is represented as <query-id, query-text, item-ids>. Relevant parameter settings during the pretraining of the CLIP model are presented in Table~\ref{tab1}.

\begin{table}[thb]
  \small
  \centering
  \caption{Pre-training CLIP Parameters Settings}
    \begin{tabular}{cc|cc}
    \toprule
    \textbf{Parameters}&\textbf{Value}&\textbf{Parameters}&\textbf{Value} \\
    \midrule
    context\_length & 52    & epoch & 25 \\
    batch\_size(i/t) & 64/64 & num\_workers & 4 \\
    learning\_rate & 3e-06 & distillation & False \\
    \bottomrule
    \end{tabular}
  \label{tab1}
\end{table}

\begin{figure}[thb]
  \centering
  \includegraphics[width=1\linewidth]{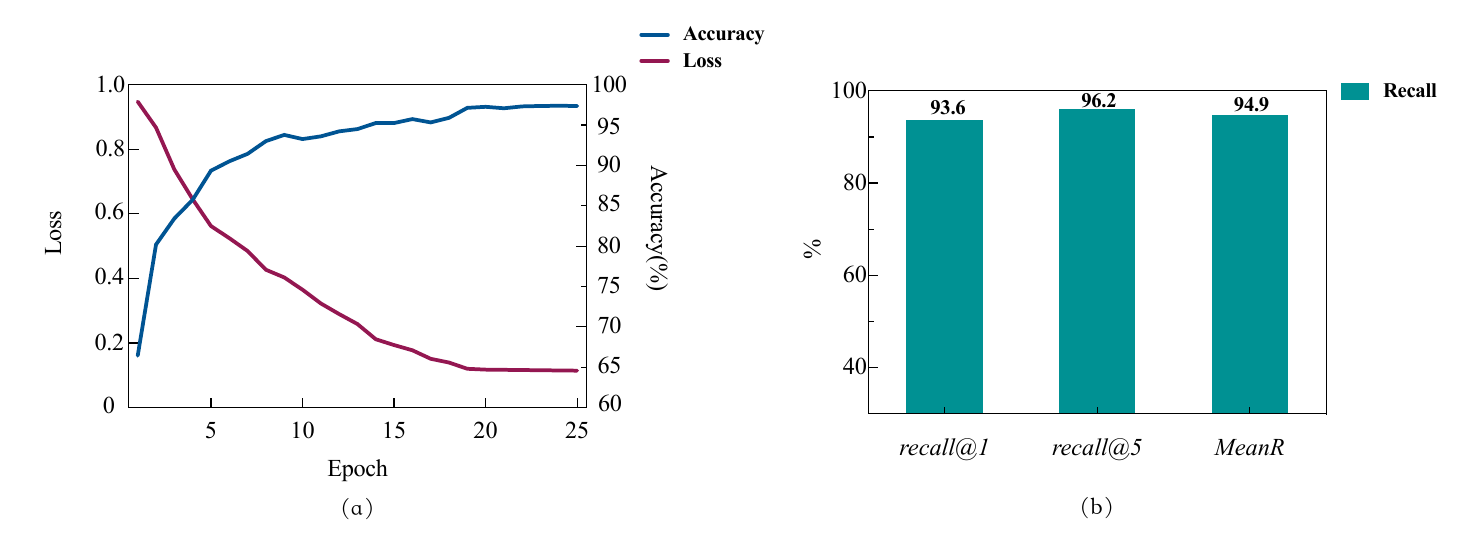}
  \caption{The performance metrics of CLIP on the MUGE dataset are presented as follows: (\textit{a}) The trend of loss and accuracy during training is depicted in the chart. (\textit{b}) The chart illustrates the recall rates and average recall rates of CLIP models at $k$=1 and $k$=5.}
  \label{fig.4}
\end{figure}

\textbf{Fastspeech2} Utilizing the Baker~\footnote{https://www.data-baker.com/data} Chinese female speech dataset, the mel-spectrogram generator and the variable adapter in the Fastspeech2 vocoder are pre-trained to synthesize high-quality audio. The Baker dataset comprises 10,000 text-speech pairs, with audio sampled at a frequency of 48 kHz with 16 bits. The average word length is 16 words, and the dataset comprehensively covers phonetics, types, tones, sound connections, and rhythm. The relevant parameter settings for pre-training the Fastspeech2 model are outlined in Table~\ref{tab2}.

\begin{table}[thb]
  \small
  \centering
  \caption{Pre-training Fastspeech2 Parameters Settings}
    \begin{tabular}{cc|cc}
    \toprule
    \textbf{Parameters} & \textbf{Value} & \textbf{Parameters} & \textbf{Value}\\
    \midrule
    sampling\_rate & 16000    & epoch & 100 \\
    filter\_length & 1024 & mel\_fmin & 0 \\
    hop\_length & 256 & mel\_fmax & 8000 \\
    max\_wav & 32786 & mel\_channels & 80 \\
    \bottomrule
    \end{tabular}%
  \label{tab2}%
\end{table}%

\begin{figure*}[thb]
  \includegraphics[width=1\linewidth]{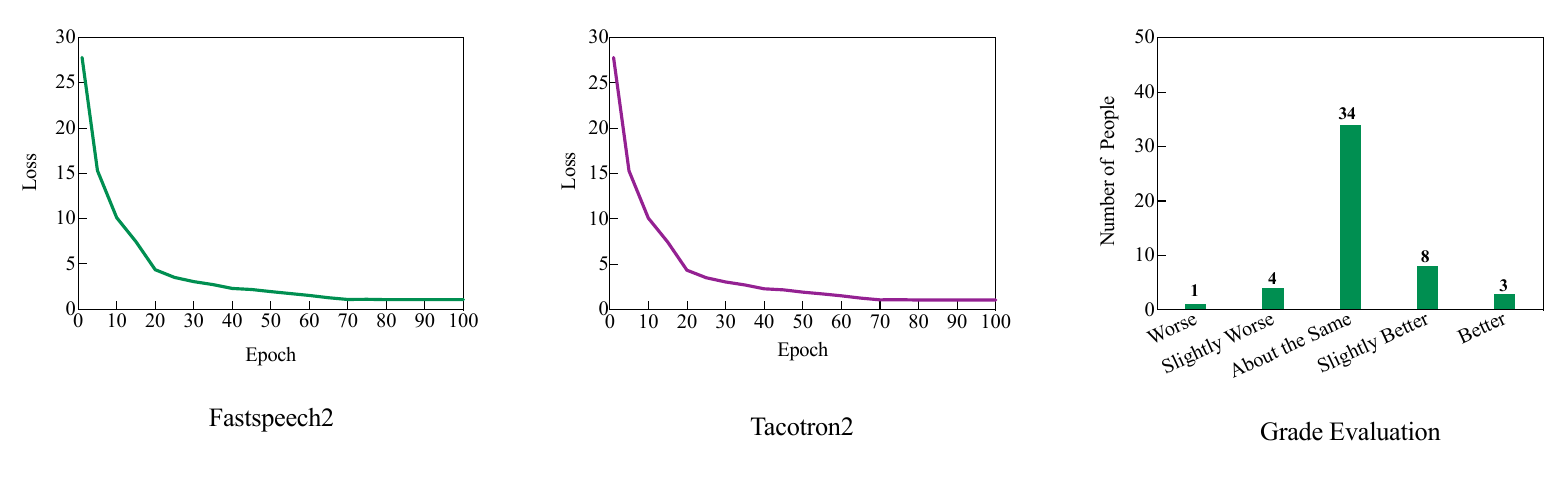}
  \caption{The performance on the Baker dataset is documented as follows:(\textit{a}) The trend of loss variation during model training is recorded, with significant fluctuations observed in the first 5 epochs followed by minimal changes in subsequent epochs.(\textit{b}) The distribution of scores from 50 participants, who rated the naturalness of the synthesized speech generated by Fastspeech2, is presented across five rating levels.}
  \label{fig.5}
\end{figure*}

According to the pre-training parameters of CLIP and Fastspeech2 in Table 1 and Table 2 and the evaluation indicators in Section 4.1, the convergence and performance of the two models can be clearly observed in Figure~\ref{fig.4} and Figure~\ref{fig.5}

From Figure~\ref{fig.4}, it can be observed that under the set number of training iterations, the training loss and accuracy of CLIP models tend to stabilize. Due to the ability of the CLIP$_{ViT-H/14}$ model to handle higher resolution images, it achieves higher accuracy in extracting features from image-text pairs. The recall rate of the KNN network retrieval also gradually improves with the feature extraction capabilities of the visual encoder.The average recall rate of the CLIP$_{ViT-H/14}$ model reached 94.9\%, achieving a good pre-training effect.

In Figure~\ref{fig.5}, the convergence of the loss values of the Fastspeech2 model after 100 iterations of training is recorded. Subsequently, the number of evaluators for each rating category is tallied to calculate the \textit{MOS} value for the synthesized speech, and the number of mispronounced phonemes in each audio is computed to quantify the audio quality. According to the equations (14) and (15), the \textit{MOS} value is calculated to be 0.16, indicating that the naturalness of the synthesized speech generated by the Fastspeech2 model aligns with normal language levels. The calculated \textit{WER} value of 4.2\% also falls within a relatively low range. Considering both \textit{MOS} and \textit{WER} metrics, it can be inferred that the performance of the Fastspeech2 model in synthesizing speech significantly improves after pretraining, rendering it suitable for subsequent fine-tuning in specific domains.

\section{Experiments}
In order to analyze the performance of the model CLIP-KNN-Fastspeech2 and the effect of the strategy of independent pre-training followed by joint fine-tuning, we designed two experimental schemes: 

(1) \textbf{\textit{Compare with the baseline models}}\quad From the perspectives of subjective feelings and objective indicators, we evaluate the performance of this model and the existing SOTA model on three public datasets Flickr8k,VGGSound and ImageHear.

(2) \textbf{\textit{Compare with other deep learning models}}\quad \textit{Different training strategies}: the two-stage output (image-text-audio) after independent pre-training VS the end-to-end output (image-audio) after pre-training and joint fine-tuning,\textit{different I2T models} and \textit{different T2A models}.

Note that in order to verify experiment (2), we used a Chinese public dataset Wukong and a self\-built dataset BIT-DP.

(3) \textbf{\textit{The impact of different image-speech data ratios}}\quad We set different image-audio data ratios for training on the datasets VGGSound and Wukong. Compare the changes in indicators between our model and the suboptimal models in Table 4 and 6.

\subsection{Metrics}
In order to save the training time and the data volume requirements of the deep neural network, we use the strategy of pre-training CLIP and Fastspeech2 modules independently, and then jointly fine-tuning.

The two-stage model proposed in this paper first pretrains the CLIP and Fastspeech2 models separately, and finally joint fine-tunes them on the self-built BIT dataset. In the I2T task stage, the objective evaluation metric accuracy (\textit{ACC}) will be used to measure the effectiveness of the model's image-text feature matching. Then, metrics such as \textit{Recall@1}, \textit{Recall@5}, and \textit{MeanR} are set to evaluate the KNN network's image-text retrieval capability. Specifically, refer to Equations (11) to (13):
\begin{equation}
  \ ACC=\frac{TP+TN}{TP+TN+FP+FN}  
\end{equation}
\begin{equation}
  \ Recall@k=\frac{TP@k}{TP@k+TN@k} 
\end{equation}
\begin{equation}
  \ MeanR=\frac{\sum_{i=1}^{T}\left (   Recall@k\right )_{i}}{T} 
\end{equation}
In which, $TP$ and $FP$ represent correctly and incorrectly predicted positive examples, while $TN$ and $FN$ represent correctly and incorrectly predicted negative examples, respectively. $MeanR$ stands for the average recall across multiple $k$ values.

In the T2A task stage, the subjective qualitative metric \textit{MOS} score (Mean Opinion Score) is improved, with five levels [\textit{Worse, Slightly Worse, About the Same, Slightly Better, Better}] set, and 20 individuals of different genders, ages, and educational backgrounds are invited to subjectively evaluate the naturalness of synthesized audio. Meanwhile, the objective evaluation metric \textit{WER} is introduced to measure the pronunciation accuracy of the Fastspeech2 model in synthesizing audio. It is noteworthy that the \textit{WER} metric is calculated by manually judging the pronunciation of 20 randomly synthesized sentences. Shown as  equations (14) and (15) for details:
\begin{equation}
  \ MOS=\frac{\sum_{j=1}^{5} S_{j} H_{j}}{H} 
\end{equation}
\begin{equation}
  \ WER=\frac{\sum_{K=1}^{N}W_{K} /W_{K}^{t}  }{N}  
\end{equation}

where $H$ represents the total number of participants in the scoring process; $S_{j}$ denotes the score corresponding to the $j$-th level, with $S_{j}$=[1,2,3,4,5];where a value closer to 5 indicates higher naturalness of the generated speech, and vice versa; $H_{j}$ represents the number of participants who scored $S_{j}$. In equation (15), $N$ denotes the total number of audio samples to be evaluated, $W_{K}$ represents the number of incorrectly pronounced characters in the $k$-th audio sample, and $W_{K}^t$ represents the total number of characters in the $k$-th audio sample.

In the Comparative Experiments , in addition to the \textit{WER} metrics, the \textit{MCD} ~\citep{kubichek1993mel} metric is introduced to evaluate the ability of the CLIP-KNN-Fastspeech2 model to generate speech from images. \textit{MCD} measures the quality of synthesized speech by calculating the distance between the Mel cepstral sequences of the original and synthesized audio. Here, the original audio is Chinese speech synthesized by the Coqui-TTS\footnote{https://github.com/coqui-ai/TTS} speech synthesis tool, while the speech generated by our model serves as the test audio. It is noteworthy that both the \textit{WER} and \textit{MCD} metrics are calculated based on a random selection of 20 synthesized speech audio samples.At the same time, a Chinese speech recognition system is directly invoked\footnote{https://github.com/nl8590687/ASRT\_SpeechRecognition}, using voice as input and pinyin or text as output. The project model was used to generate a text transcription of the corresponding speech, which was compared with the real text description from the dataset Flickr-8k~\citep{rashtchian2010collecting} .This makes it easier for us to use the BLEU4~\citep{papineni2002bleu} and METEOR~\citep{banerjee2005meteor} to evaluate the Image-to-Audio model in different language scenarios.Meanwhlie,\textit{CS(CLIP-Score)}~\citep{sheffer2023hear} and \textit{FAD}(Fr´echet Audio Distance)~\citep{kilgour2018fr}are also used to further increase the credibility of the performance comparison with the SOTA model.

\subsection{Comparative Experiments}

\begin{table*}[thb]
  \centering
  \caption{Objective Comparison with SOTA Methods on Flickr8k,VGGSound and ImageHear (word-level comparison)}
  \resizebox{\linewidth}{!}{
    \begin{tabular}{cccccc}
    \toprule
    \textbf{DATASET} & \textbf{MODEL} &\textbf{BLEU4}$\uparrow$ &\textbf{METEOR}$\uparrow$ &\textbf{CS}$\uparrow$ & \textbf{FAD}$\downarrow$\\
    \midrule
      &SAT\cite{hsu2020text}                & 11.6  &  14.1  &$-$&$-$ \\
      &SAT-FT\cite{hsu2020text}             & 12.5  &  14.5 &$-$&$-$ \\
      Flickr8k  &Image2speech\cite{effendi2021end}    & 14.8 &  17.4 &$-$&$-$ \\
      &CLIPSonic-IQ\cite{dong2023clipsonic}    & $-$ &  $-$ &7.743& 3.754 \\
      &Im2Wav\cite{sheffer2023hear}    &$-$ & $-$ & \textbf{8.685} &8.022\\
      &Our Model                             & \textbf{15.0} & \textbf{17.6}
      &8.442 &\textbf{2.902}\\\midrule
      &CLIPSonic-IQ\cite{dong2023clipsonic}    &$-$& $-$& 7.251& 3.495 \\
      VGGSound&Im2Wav\cite{sheffer2023hear}    & $-$&  $-$& 7.827& 6.005\\
      &Our Model                             & 16.2 & 16.5&\textbf{9.033} &\textbf{2.377}\\
    \midrule
      &CLIPSonic-IQ\cite{dong2023clipsonic}    & $-$ &  $-$ & 11.392& $-$\\
      ImageHear&Im2Wav\cite{sheffer2023hear}    & $-$ &  $-$ & 9.843& $-$\\
      &Our Model                             & 17.7 & 18.2
      &\textbf{13.032} &$-$\\
    \bottomrule
    \end{tabular}
    }
  \label{tab11}
\end{table*}

The experimental operating system in this article is Linux Ubuntu 20.04, the video memory is 40GB, and the GPU is A100. The Python version is 3.8, the pytorch version is 1.11.0.

\subsubsection{Compare with the Baseline Model}

As shown in Table~\ref{tab11} and Table~\ref{tab13}. We selected the following baseline models for experiments:SAT~\citep{hsu2020text},SAT-FT~\citep{hsu2020text},Image2speech~\citep{effendi2021end},
CLIPSonic-IQ~\citep{dong2023clipsonic} and Im2Wav~\citep{sheffer2023hear}.

\begin{table*}[htbp]
  \centering
  \caption{Subjective Comparison with SOTA Methods on Flickr8k,VGGSound and ImageHear.(\textbf{MOS})}
\resizebox{\linewidth}{!}{
  \begin{tabular}{ccccc}
\toprule
\textbf{MODEL} & \multicolumn{2}{c}{\textbf{VGGSound}}                                                        & \multicolumn{2}{c}{\textbf{ImageHear}}                                                       \\ \cline{2-5} 
                       & Fidelity$\uparrow$                                 & Relevance $\uparrow$                               & Fidelity  $\uparrow$                               & Relevance $\uparrow$                               \\ \midrule
CLIPSonic-IQ\cite{dong2023clipsonic}           & 1.838±0.511                              & 2.415±0.645                              & 1.840±0.502                              & 2.705±.398                               \\
Im2Wav\cite{sheffer2023hear}                 & 2.533±0.522                              & 2.140±0.551                              & 2.888±0.502                              & 3.215±0.291                              \\
Our Model              & \multicolumn{1}{l}{\textbf{2.720±0.460}} & \multicolumn{1}{l}{\textbf{2.700±0.680}} & \multicolumn{1}{l}{\textbf{3.020±0.620}} & \multicolumn{1}{l}{\textbf{3.240±0.340}} \\ \bottomrule
\end{tabular}
 }
\label{tab13}%
\end{table*}

\begin{table*}[thb]
\caption{Comparison of Different DL Models on Wukong and BIT-DP}
\resizebox{\linewidth}{!}{
\begin{tabular}{cccccc}
\toprule
\textbf{NUMBER} & \textbf{MODEL} & \multicolumn{2}{c}{\textbf{Wukong}}    & \multicolumn{2}{c}{\textbf{BIT-DP}}    \\ 
\cline{3-6} 
                        &                        & WER($\%$)$\downarrow$      & Mean\_MCD$\downarrow$      & WER($\%$)$\downarrow$      & Mean\_MCD$\downarrow$      \\ 
\midrule
1                       & CNN-LSTM+Tacotron2     & 9.5          & 8.103          & 9.7          & 8.332          \\
2                       & VLBERT+Tacotron2       & 7.2          & 6.062          & 7.6          & 6.437          \\
3                       & VLBERT+Fastspeech2     & 5.8          & 4.877          & 6.4          & 5.282          \\
4                       & CLIP-KNN+Fastspeech2   & 4.9          & 4.228          & 5.1          & 4.708          \\
5                       & Our Model              & \textbf{4.1} & \textbf{3.276} & \textbf{4.2} & \textbf{3.622} \\ 
\bottomrule
\end{tabular}
}
\label{tab9}
\end{table*}

\subsubsection{Compare with Other DL Models} 
The I2T model includes CNN-LSTM\cite{xu2015show}, VLBERT (Visual-Linguistic BERT)~\citep{su2019vl} and CLIP, and the T2A model mainly uses Tacotron2 and Fastspeech2.

\textbf{CNN-LSTM}\quad The attention mechanism is mainly used to correspond the image block features extracted by CNN to the words predicted by LSTM.

\textbf{VLBERT}\quad A Visual Feature Embedding layer is added to the input of BERT to embed the image features extracted by Fast R-CNN. Then, the image and text are used as clues for mask prediction.

\textbf{Tacotron 2}\quad WaveNet replaces the Griffin-Lim algorithm in Tacotron and uses LSTM and convolutional layers to improve the similarity of synthesized speech waveforms to human voices.

For more information about model CLIP and Fastspeech2, please refer to section 3.4 and will not be repeated here.

From Table~\ref{tab9}, the following conclusions can be drawn:

\textbf{(1)} The experimental group with different I2T models was composed of model 1 and 2, model 3 and 4. The comparison of the two experimental groups shows that the VLBERT model is better than the CNN-LSTM for image and text retrieval, which decreases by 2.1\% on the \textit{WER} index and nearly two percentage points on the \textit{Mean\_MCD} index.In the same way, the CLIP-KNN model is easier to capture the feature relationship between images and texts than the VLBERT model.

\textbf{(2)} Model 2 and Model 3 were composed of experimental groups using different T2A models. The Fastspeech2 model is better than the Tacotron2 model in terms of \textit{WER} and \textit{Mean\_MCD}, and Fastspeech2 shows better performance in terms of the accuracy of audio synthesis.

\textbf{(3)} Experimental groups with different output forms were composed of model 4 and model 5. Here, "+" means that the I2T model is used to output the text first, and then the text is processed into the input of the subsequent T2A model to synthesize speech. From the changes in the \textit{WER} and \textit{Mean\_MCD} values of the two indicators, we can clearly see the strategic advantages of integrating multiple basic models.

In view of the above conclusions, the explanation is as follows:

\textbf{(1)} CNN-LSTM uses CNN to extract useful spatial features from the image, and transforms these features into a feature vector containing the overall information and local details of the image, uses the LSTM to capture the long dependencies in the text, and then realizes the matching of the image feature vector and the text feature vector in the semantic space through the attention mechanism. However, the use of LSTM can only focus on the current position and the next word position when processing long texts, resulting in the forgetting of semantic information between contexts. Through the stacking of multi-layer multi-modal Transformer attention modules, the VLBERT model can derive representations with rich visual language cue aggregation and alignment functions. Compared with VLBERT's efficient collaboration in dealing with cross-modal tasks, CLIP establishes a strong connection in cross-modal semantics and can maintain high feature recognition ability in the case of drastic domain changes.

\textbf{(2)} Compared with the Tacotron2 model, which controls the quality of synthesized speech by integrating the prediction loss, binary classification loss, and the loss of the Mel spectrum generated by the postnet network and the target Mel spectrum, Fastspeech2 optimizes and reduces the loss of audio and energy that affect the speech quality to generate audio that is more similar to human language.

\textbf{(3)} Compared with the conversion output of image-text-audio modality, the strategy of fine-tuning and training multiple basic models at the same time by designing the target loss takes into account the capabilities of I2T and T2A models and minimizes the modal conversion loss.
\begin{figure}[ht]
  \includegraphics[width=0.85\linewidth]{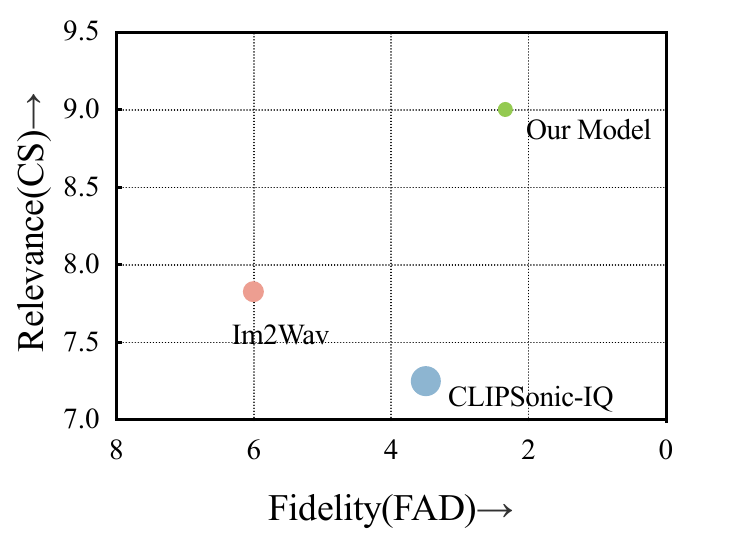}
  \caption{Our method achieves better results on both metrics and possesses faster inference speed(samller circle size).}
  \label{fig.10} 
\end{figure}
From the comparison of the above experimental results, it can be concluded that the synthetic sound quality effect of the proposed model shows certain advantages in objective analysis, perceptual cognition and model reasoning speed compared with the existing models.

\subsubsection{The impact of different image-speech data ratios} Setting different training sample ratios on the VGGSound dataset and compared the changes in the \textbf{\textit{CS}} and \textbf{\textit{FAD}} indicators of the CLIPSonic-IQ~\citep{dong2023clipsonic} model. Similarly, we compared the model in this paper with the VLBERT+Fastspeech2 model on the Wukong dataset and observed the changes in \textbf{\textit{WER}} and \textbf{\textit{Mean\_MCD}}.

\begin{figure*}[thb]
  \includegraphics[width=1\linewidth]{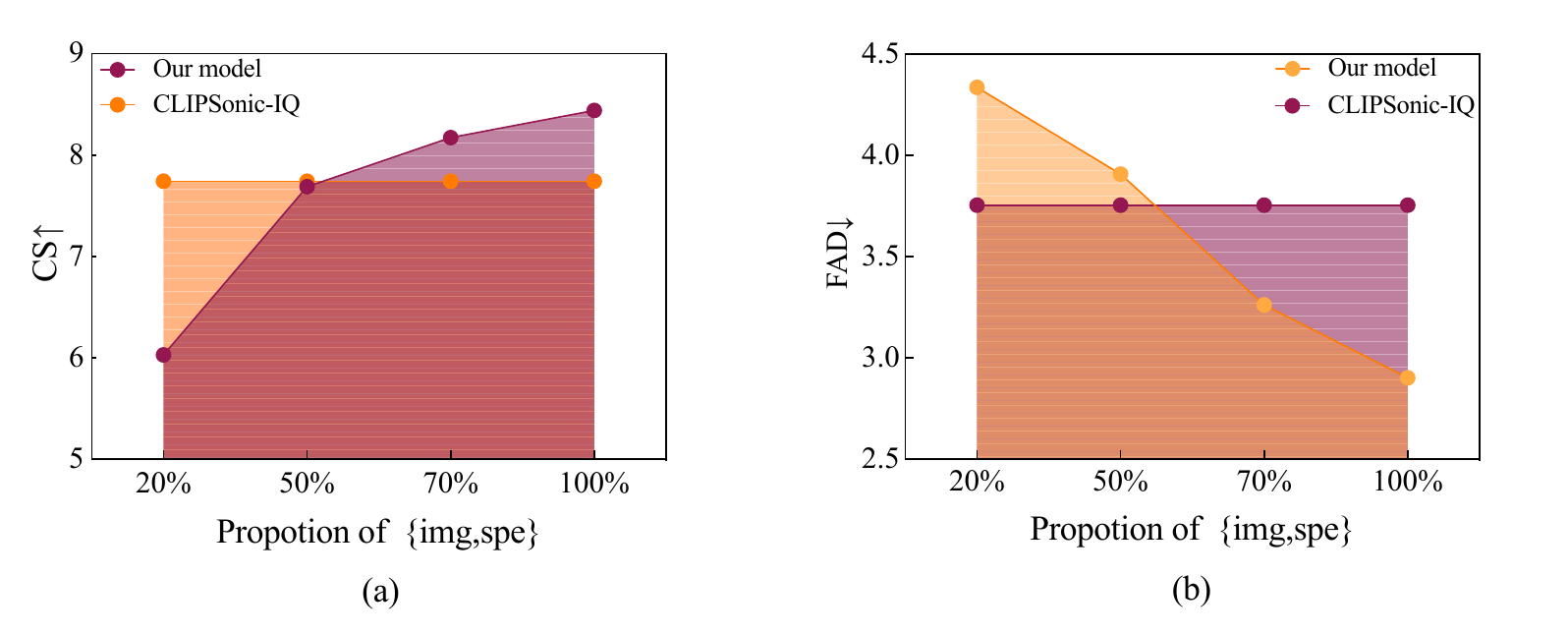}
  \caption{The numerical changes of CS and FAD on VGGSound with different training set ratios.}
  \label{fig.11n}
\end{figure*}

\begin{figure*}[thb]
  \includegraphics[width=1\linewidth]{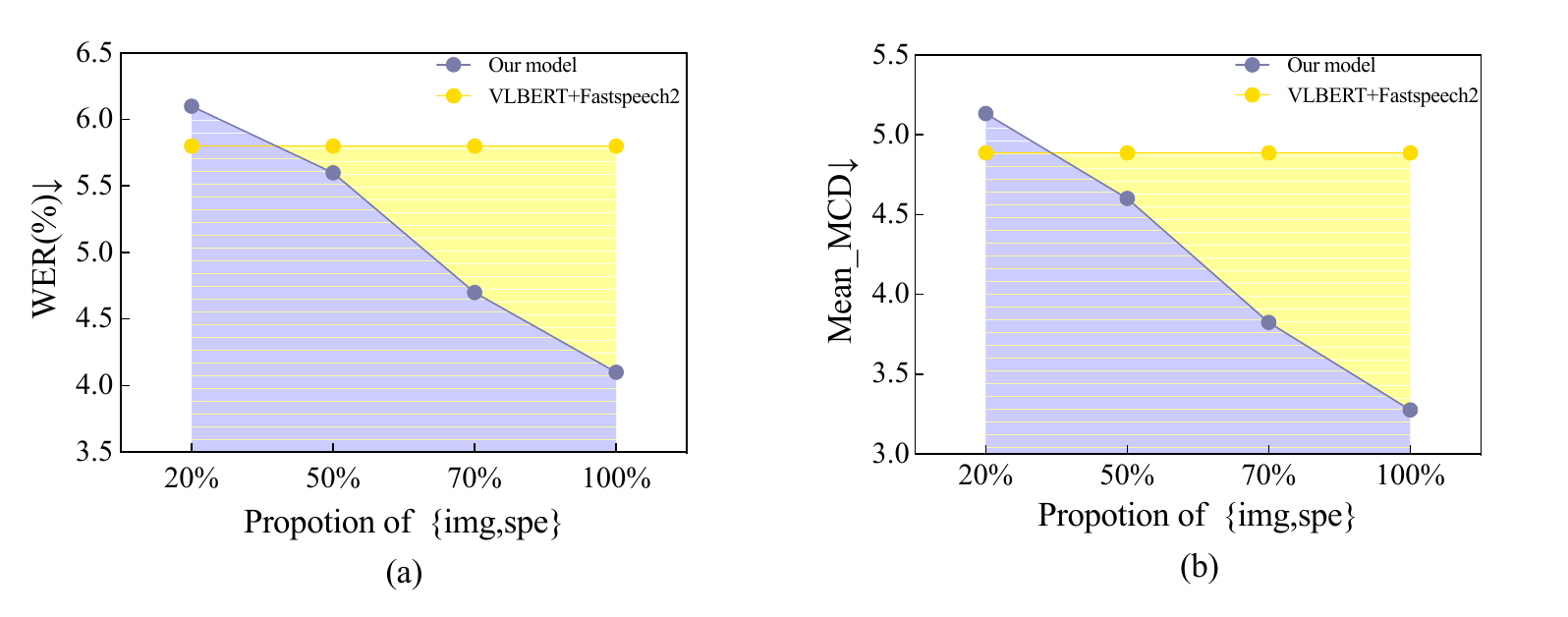}
  \caption{The numerical changes of WER and Mean\_MCD on Wukong with different training set ratios.}
  \label{fig.12n}
\end{figure*}

It should be noted that the training set size is considered to be strongly positively correlated with performance. Therefore, we only compared the best performance of the suboptimal model (only discussing the indicator values of the suboptimal model when the training set size is 100\%).

From the visualization results in Figure~\ref{fig.11n},~\ref{fig.12n}, we can see that our model can still achieve similar index values to the suboptimal model when reducing the training samples by 50$\%$-55$\%$.

\subsection{Ablation Experiments}
Through the above comparative experiments, the model CLIP-KNN-Fastspeech2 shows good speech synthesis effect. In order to further analyze the influence of the basic model on the effect of synthetic speech, the self-built BIT-DP dataset was used to analyze the effect of the CLIP model using different visual encoders or different vocoder combinations on braille image-speech conversion (80\% in the training set and 20 $\%$ in the test set).

\subsubsection{Convergence Analysis}
\begin{figure*}[thb]
  \includegraphics[width=1\linewidth]{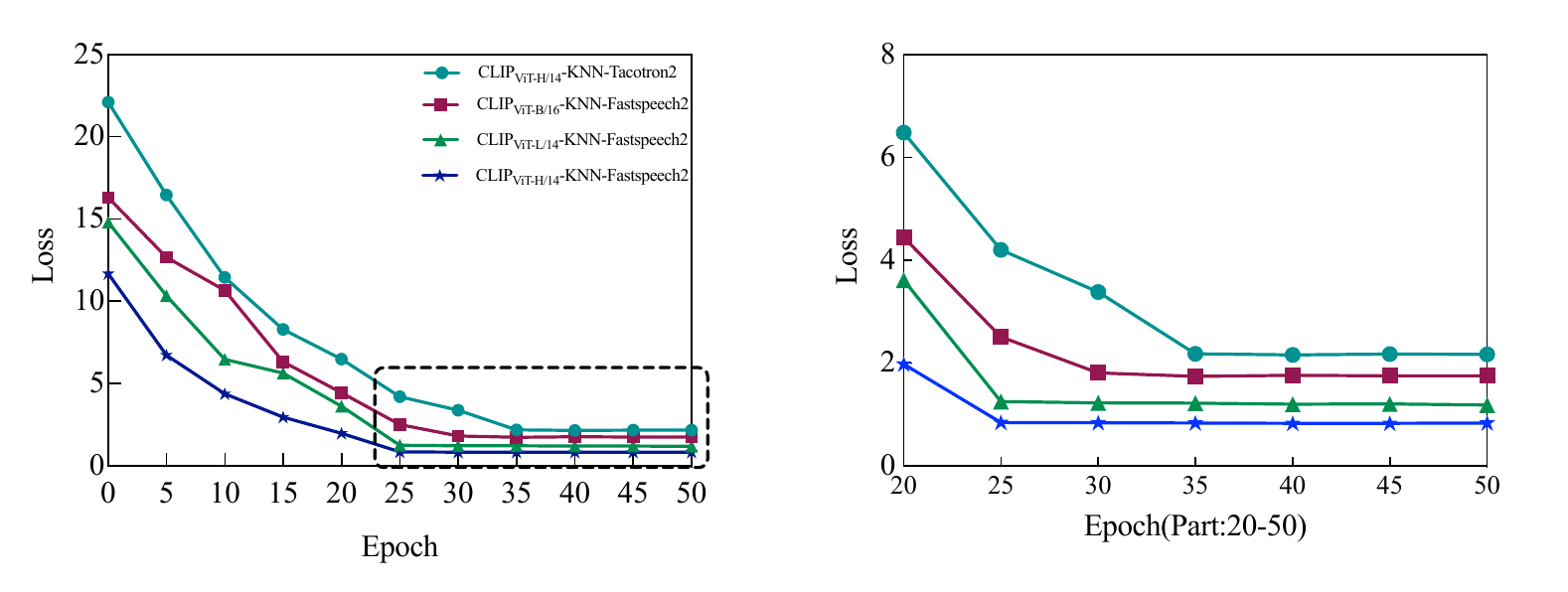}
  \caption{The convergence situation of model fine-tuning on the BIT-DP dataset.}
  \label{fig.6}
\end{figure*}

The modality conversion capabilities in the I2T and T2A stages directly influence the quality of the output speech. Due to variations in feature extraction capabilities among different visual encoders for braille images and differences in optimization functions for audio factors such as energy and pitch among different speech synthesizers, a comparison of loss convergence among different models will be conducted on the same BIT-DP dataset (epoch=50), as depicted in Figure~\ref{fig.6}.

The key difference lie in Figure~\ref{fig.5} is that Figure~\ref{fig.6} depicts more iteration rounds are required for individual vocoder pre-training. After fine-tuning, the loss trends of the four models stabilize and converge after 50 epochs, indicating that the efficiency of step-wise pre-training followed by joint fine-tuning is higher than directly joint training. From the overall loss trend graph, it can be observed that the model CLIP$_{ViT-H/14}$-KNN-Fastspeech2 has the fastest convergence speed and the smallest convergence loss, while the model CLIP$_{ViT-H/14}$-KNN-Tacotron2 has the slowest convergence speed and the largest convergence loss. Zooming in on the details from epoch 20 to 50: 

\textbf{(1)} The model CLIP${ViT-H/14}$-KNN-Fastspeech2 starts to converge after 25 iterations,which is 10 epochs earlier than the model CLIP$_{ViT-H/14}$-KNN-Tacotron2.
\textbf{(2)} The convergence time points and convergence loss values of the models CLIP$_{ViT-L/14}$-KNN-Fastspeech2 and CLIP$_{ViT-H/14}$-KNN-Fastspeech2 are consistent, while they differ from the convergence loss of the model CLIP$_{ViT-B/16}$-KNN-Fastspeech2.

Analysis of the above results:

\textbf{(1)}The different model architectures involved only vary in the TTS (Text to Speech) model. Fastspeech2, based on the Transformer architecture, updates parameters in parallel layers, unlike the Tacotron2 model, which employs an LSTM (Long short-term memory) decoder. In Tacotron2, the generation of the mel-spectrogram at time t depends on the output of the previous time step, which significantly impacts the convergence speed between the two models.

\textbf{(2)}The main difference in the models involved lies in the visual encoder. Although ViT-B/16, ViT-L/14, and ViT-H/14 all utilize the Transformer architecture, the latter two have 16 attention heads and adopt a patch size of 14×14 pixels. Therefore, when processing the same image, the latter two are more likely to highlight image details and capture important local features compared to ViT-B/16, which has 14 attention heads and a patch size of 16×16 pixels.

\subsubsection{Audio Quality Analysis}
After analyzing the loss convergence trend of the model, we introduce \textit{MOS} indicators to quantitatively analyze the human subjective perception of the synthesized speech of different visual encoder models.Firstly, the synthesized speech was manually rated, and the number of ratings for different levels of speech naturalness was counted. Then, we tested the significance of the obtained voting results with a 95\% confidence interval and obtained a $P_i$<0.05, where $i$ represents the model number. Prove that the voting results are representative, and of course we have also introduced the basic information of the people who participated in the voting in the Table~\ref{tab12}.
\begin{table}[thb]
  \centering
  \caption{General Information of Voters.
  $\mu$:AVERAGE,$\sigma$:STANDARD DEVIATION,HE: with HIGHER EDUCATION}
    \resizebox{1\linewidth}{!}{ 
    \begin{tabular}{ccc}
    \toprule
    \textbf{INFORMATION} & \textbf{MALE}& \textbf{FEMALE} \\
    \midrule
    Number of subjects     &  28 & 22\\
    Age[years]($\mu$±$\sigma$)    &  26±4.7&  24±3.1 \\
    Range of age[years]    & 20-33 & 19-29 \\
    Education background     & 79$\%$(HE)21$\%$(other) & 82$\%$(HE)18$\%$(other)\\
    \bottomrule
    \end{tabular}
    }
  \label{tab12}
\end{table}

\begin{figure}[thb]
  \centering
  \includegraphics[width=0.9\linewidth]{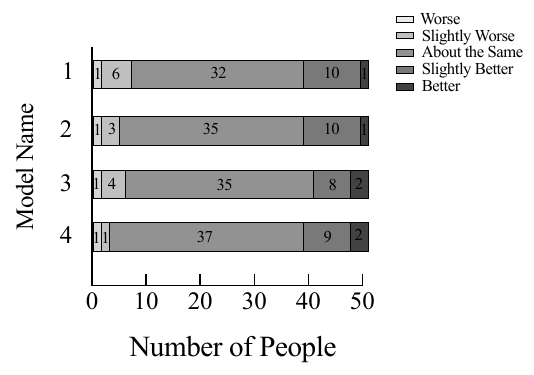}
  \caption{Count of Ratings (The numbers in the figure represent models 1: CLIP$_{ViT-H/14}$-KNN-Tacotron2, 2: CLIP$_{ViT-B/16}$-KNN-Fastspeech2, 3: CLIP$_{ViT-L/14}$-KNN-Fastspeech2, 4: CLIP$_{ViT-H/14}$-KNN-Fastspeech2, and the numbers in the figure represent the number of ratings for each level.)}
  \label{fig.7}
\end{figure}

In Figure~\ref{fig.7}, it can be observed that the varying shades of color correspond to the degree of naturalness of the speech. The distribution of the number of ratings for each level of speech naturalness generally follows a normal distribution pattern, consistent with the statistical principles of rating scales. Subsequently, when calculating the \textit{MCD} metric, the standard mode is used to compute the direct Euclidean distance based on Mel Cepstral coefficients. Finally, the mean value of \textit{MCD} for 20 speech samples is calculated and denoted as \textit{Mean\_MCD}. The specific results of the metric calculations are shown in Table~\ref{tab4}.
\begin{table}[thb]
  \centering
  \caption{Audio Quality Analysis with Different Visual Encoders}
  \resizebox{\linewidth}{!}{
    \begin{tabular}{cccc}
    \toprule
    \textbf{NUMBER} &  \textbf{Fidelity(MOS)}$\uparrow$ &\textbf{WER}($\%$)$\downarrow$&\textbf{Mean\_MCD}$\downarrow$ \\
    \midrule
    1     &  2.36  & 6.2   & 5.208 \\
    2     &  3.14  & 4.9   & 4.512 \\
    3     &  3.08  & 4.4   & 3.934 \\
    4     &  \textbf{3.2} & \textbf{4.2} & \textbf{3.622} \\
    \bottomrule
    \end{tabular}
  }
  \label{tab4}
\end{table}
Conclusions drawn from Table~\ref{tab4} are as follows:

\textbf{(1)}Comparing the performance of the four models across three metrics reveals that Model 4 exhibits the most prominent ability in the image-to-speech modal conversion.

\textbf{(2)}Comparing Model 1 and 4, it is evident that Model 4 outperforms Model 1 in both naturalness \textit{MOS} and \textit{WER} metrics. This may be attributed to the introduction of variable adapters in the TTS model Fastspeech2, which adjusts the naturalness of speech by optimizing the loss functions for phoneme duration, energy, and pitch. In contrast, the Tacotron2 model synthesizes speech based on the generated mel-spectrogram using a modified WaveNet. Regarding the \textit{MCD} metric, Model 4 also outperforms Model 1. Unlike Model 1, which only utilizes a 5-layer CNN network as Post-net to adjust the generated mel-spectrogram, Model 4 simultaneously considers the generation of the mel-spectrogram and the alignment of the mel-spectrogram with the text embeddings in the position encoding (time frames).

\textbf{(3)}Comparing Model2, Model3, and Model4, it is observed that the pixel size sampled by the visual encoder and the number of attention heads have an impact on speech synthesis under the same TTS model. However, the difference in \textit{WER} and \textit{Mean\_MCD} metric values is not significant, possibly due to the limited image size and local features of the BIT dataset. (\textbf{1.} Average image size: 120×80, \textbf{2.} Features: point position information and point count information).

\subsection{Futher Analysis}
In this section, we will analyze the following data and loss weight parameters from two perspectives:\textbf{Given the inconsistent number of braille cells} and \textbf{composition rules among different data types}, the potential data noise also has a negative effect on the model performance. To further analyze the impact of data types and data processing on model performance, suitable datasets were selected for experimentation according to the dataset division criteria outlined in Section 4.3.At the same time, the weight parameter combinations lost at different stages were designed to explore the influence of the conversion ability of each stage on the final speech synthesis effect.

\subsubsection{the Impact of Data Types}
\begin{figure*}[thb]
  \centering
  \includegraphics[width=\linewidth]{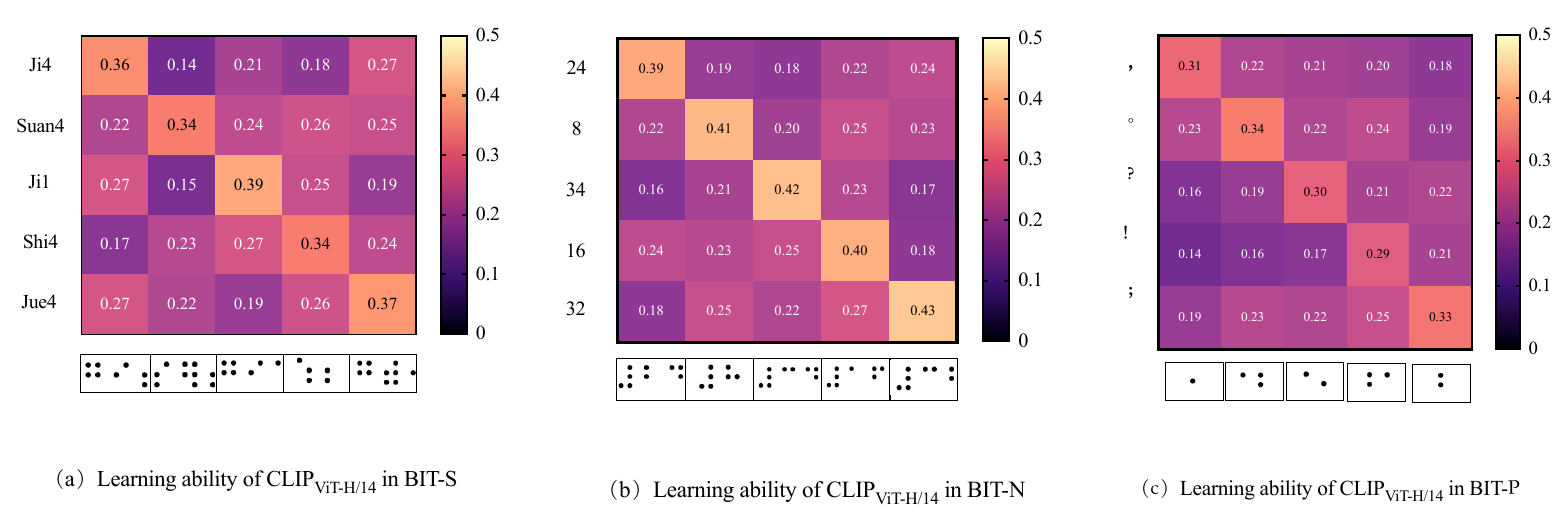}
  \caption{Cosine similarity of image-text pairs (Unnormalized)}
  \label{fig.8}
\end{figure*}

Using the I2T model CLIP$_{ViT-H/14}$, experiments were conducted on datasets BIT-N, BIT-S, and BIT-P to assess the model's ability to learn features for the three types of data: numbers, pinyin, and punctuation. The cosine similarity of image-text matching was employed for this evaluation, as illustrated in Figure~\ref{fig.8}.

In Figure~\ref{fig.8}, the cosine similarity of image-text pairs for three types of data is computed to analyze the capability of model CLIP$_{ViT-H/14}$ in extracting features from both images and texts. Here, the average of the sum of cosine similarity ranges for each column is calculated to quantify which type of data the model is more adept at recognizing. The average range in Figure (a) is \textbf{0.186}, in Figure (b) is \textbf{0.226}, and in Figure (c) is \textbf{0.144}. It can be concluded that the model performs best in recognizing Braille image-digit data, while it exhibits the weakest feature extraction capability for Braille image-punctuation data.

\subsubsection{the Impact of Data Processing}
In Section 4.3, two scenarios of "dirty data" and two data augmentation methods were presented. To validate the necessity of data preprocessing, experiments were conducted using the best-performing model in speech synthesis quality, CLIP$_{ViT-H/14}$-KNN-Fastspeech2, on three datasets: BIT, BIT-DC, and BIT-DA. Considering the model's runtime, 30\% of each dataset was utilized, and the \textit{WER} metric was used for quantitative comparison. The results are presented in Table~\ref{tab5}.
\begin{table}[thb]
  \centering
  \caption{Comparison of WER Values Before and After Data Processing}
  \label{tab5}
  \begin{tabular}{cc}
    \toprule
    \textbf{DATASET} & \textbf{WER}($\%$)$\downarrow$ \\
    \midrule
    BIT     &  9.1\\
    BIT-DC    &  6.4\\
    BIT-DP    & \textbf{4.2}\\
  \bottomrule
\end{tabular}
\end{table}
From the results in Table~\ref{tab5}, it is evident that the improvement in speech synthesis after data cleaning is significant. The BIT-DP dataset, derived from data augmentation based on BIT-DC after data cleaning, expands the dataset, enabling the model to extract more features related to the positions and quantities of Braille image points. Consequently, this enhancement improves the accuracy of the text and reduces the phoneme error rate in speech synthesis.

\subsubsection{Weight Parameters Analysis}
\begin{figure}[thb]
  \centering
  \includegraphics[width=\linewidth]{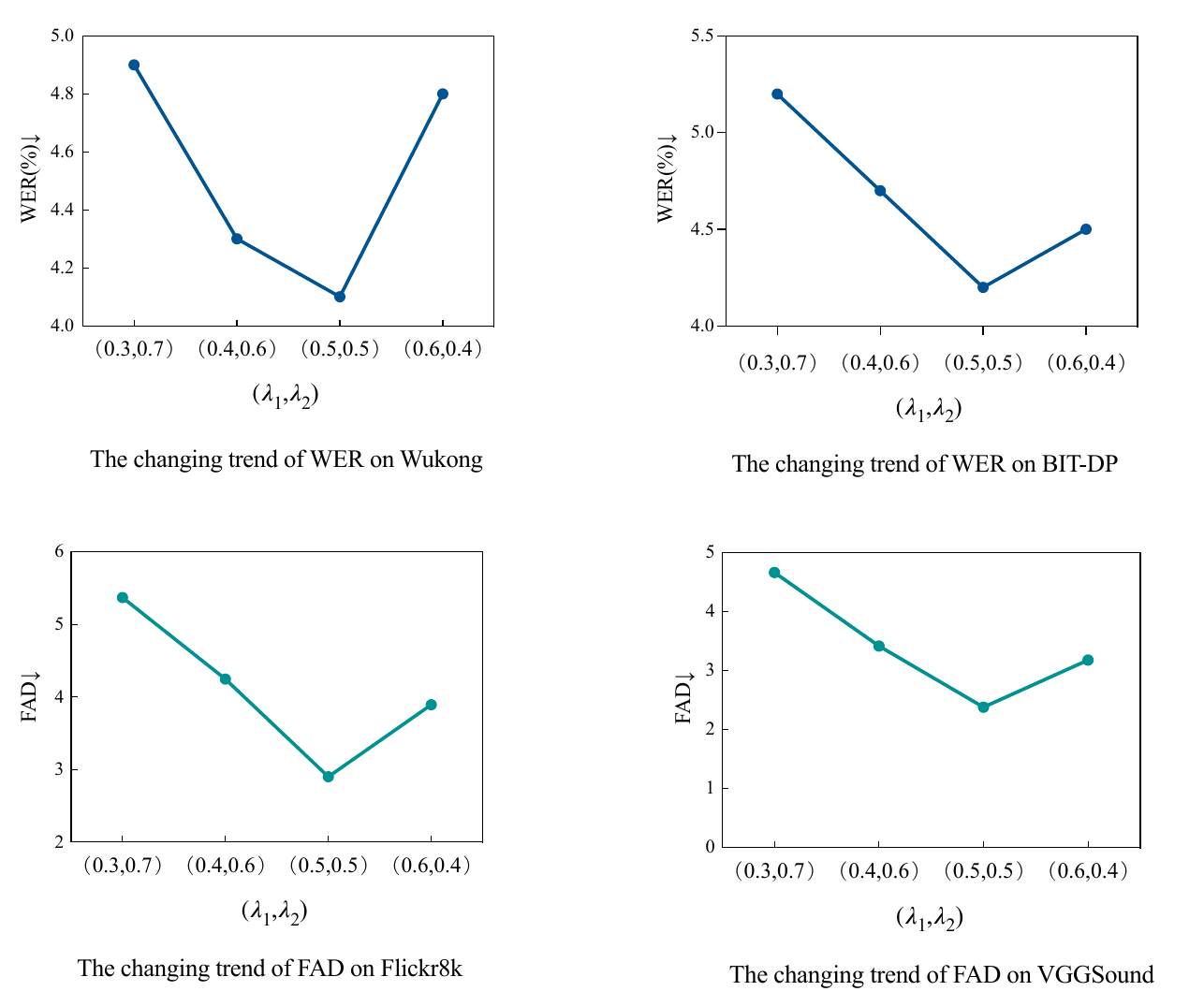}
  \caption {Comparison of WER and FAD under different combinations of weight parameters}
  \label{fig.9}
\end{figure}

According to the objective loss function $Loss_{total}$, the numerical magnitude of $\lambda_{1}$ and $\lambda_{2}$ indirectly reflects the contribution of I2T and T2A to the final speech synthesis capability in the CLIP-KNN-Fastspeech2 model. We will select the optimal model CLIP$_{ViT-H/14}$-KNN-Fastspeech2 in Table 4 for comparative analysis of \textit{WER} indicators on the test set of BIT-DP,Wukong.At the same time, we also analyzed the changes in \textit{FAD} values under different parameter combinations on the datasets VGGSound and Flickr8k. As shown in the Figure~\ref{fig.9}.
From Figure~\ref{fig.9}, our consideration of setting the initial weight parameters in Section 4 is reasonable. It is further proved that the feasibility of each component of the two-stage model designed in this paper, and the stage effect of I2T and T2A is definitely critical.

\section{Conclusion}
In this paper, we integrate basic models such as CLIP and Fastspeech2, and propose an I2A model CLIP-KNN-Fastspeech2 that facilitates visually impaired people to quickly acquire braille image knowledge. By pre-training Chinese CLIP and Chinese Fastspeech2 in parallel, and then fine-tuning on the braille image dataset. Then, by comparing the performance of multiple models on PER and other indicators, it is verified that the proposed model can achieve the output from Braille image to high-quality audio in Chinese language scenarios. At the same time, ablation experiments on CLIP models of different visual encoders were designed, and it was concluded that the visual encoder ViT-H/14 was better than the other two visual encoders in extracting image features. Finally, the recognition ability of the model for different data types (Pinyin, Punctuation, Number) is further analyzed, and the necessity of data processing is verified. In the future, the fusion ability of the basic model will be strengthened, and the semantic space module of image-text-speech will be constructed to replace the text data designed to conform to the input of the text-to-speech stage, so as to further reduce the modal conversion loss.

\bibliography{A2024_main}
\bibliographystyle{acl_natbib}

\end{document}